\pdfoutput=1

\documentclass[11pt,twoside,a4paper,cmspaper,final,collab]{cms-tdr}
\def\svnVersion{131544}\def\cmsCernNo{2012-118}\def\cmsCernDate{\today}\def\cmsMessage{Submitted to Physical Review Letters}

\begin{document}\cmsNoteHeader{BPH-12-001}

\hyphenation{had-ron-i-za-tion}
\hyphenation{cal-or-i-me-ter}
\hyphenation{de-vices}

\RCS$Revision: 131533 $
\RCS$HeadURL: svn+ssh://alverson@svn.cern.ch/reps/tdr2/papers/BPH-12-001/trunk/BPH-12-001.tex $
\RCS$Id: BPH-12-001.tex 131533 2012-06-21 17:04:15Z alverson $
\newlength\cmsFigWidth
\ifthenelse{\boolean{cms@external}}{\setlength\cmsFigWidth{0.85\columnwidth}}{\setlength\cmsFigWidth{0.4\textwidth}}
\ifthenelse{\boolean{cms@external}}{\providecommand{\cmsLeft}{top}}{\providecommand{\cmsLeft}{left}}
\ifthenelse{\boolean{cms@external}}{\providecommand{\cmsRight}{bottom}}{\providecommand{\cmsRight}{right}}
\providecommand{\PKS}{\ensuremath{\cmsSymbolFace{K}_\cmsSymbolFace{S}}\xspace}
\providecommand{\PGLz}{\ensuremath{\Lambda^{0}}\xspace}
\providecommand{\PGXb}{\ensuremath{\Xi_\cmsSymbolFace{b}}\xspace}
\providecommand{\PGXbm}{\ensuremath{\Xi^{-}_\cmsSymbolFace{b}}\xspace}
\providecommand{\PGXbz}{\ensuremath{\Xi^{0}_\cmsSymbolFace{b}}\xspace}
\providecommand{\PGXprb}{\ensuremath{\Xi^{'}_\cmsSymbolFace{b}}\xspace}
\providecommand{\PGXstb}{\ensuremath{\Xi^{\ast}_\cmsSymbolFace{b}}\xspace}
\providecommand{\PGXstbz}{\ensuremath{\Xi^{\ast 0}_\cmsSymbolFace{b}}\xspace}
\providecommand{\cPGXststb}{\ensuremath{\Xi^{\ast\ast}_\cmsSymbolFace{b}}\xspace}
\renewcommand{\cPqb}{\ensuremath{\cmsSymbolFace{b}}\xspace}
\providecommand{\PBs}{\ensuremath{\cmsSymbolFace{B}_\cmsSymbolFace{s}}\xspace}
\renewcommand{\PgLb}{\ensuremath{{\Lambda}_\cmsSymbolFace{b}}\xspace}
\newcommand{\re}{\ensuremath{\mathrm{e}}} % roman e for Euler's constant
\cmsNoteHeader{BPH-12-001} % This is over-written in the CMS environment: useful as preprint no. for export versions
	    \title{\texorpdfstring{Observation of a new $\Xi_\cmsSymbolFace{b}$ baryon}{Observation of a new Xi(b) baryon}}

\date{\today}

\abstract{
The observation of a new \cPqb baryon via its strong decay into $\Xi_\cmsSymbolFace{b}^-\pi^+$ (plus charge conjugates) is reported. The measurement uses a data sample of \Pp\Pp\ collisions at $\sqrt{s} = 7$\TeV collected by the CMS experiment at the LHC, corresponding to an integrated luminosity of 5.3\fbinv. The known $\Xi_\cmsSymbolFace{b}^-
$ baryon is reconstructed via the decay chain $\Xi_\cmsSymbolFace{b}^- \to \JPsi \, \Xi^- \to \mu^+\mu^- \, \Lambda^0 \pi^-$, with $\Lambda^0 \to \Pp \pi^-$. A peak is observed in the distribution of the difference between the mass of the $\Xi_\cmsSymbolFace{b}^-\pi^+$ system and the sum of the masses of the $\Xi_\cmsSymbolFace{b}^-$ and $\pi^+$, with a significance exceeding 5 standard deviations. The mass difference of the peak is $14.84 \pm 0.74 \, \text{(stat.)} \pm 0.28 \, \text{(syst.)}$\MeV. The new state most likely corresponds to the J$^\mathrm{P}=3/2^+$ companion of the $\Xi_\cmsSymbolFace{b}$.
}

\hypersetup{%
pdfauthor={CMS Collaboration},%
pdftitle={Observation of a new Xi(b) baryon},%
pdfsubject={CMS},%
pdfkeywords={CMS, physics, B-physics, baryon}}

\maketitle %maketitle comes after all the front information has been supplied

According to the well-established quark model and corresponding
spectroscopy of baryons, there are several predicted baryons
containing one strange and one beauty valence quark.
These include the \PGXb (ground state) and \PGXprb, both with total angular momentum and parity J$^\mathrm{P} = 1/2^+$,
a J$^\mathrm{P}=3/2^+$ state with angular momentum $\mathrm{L}=0$ (often referred to, as will be done in this Letter, as \PGXstb),
and two states with J$^\mathrm{P} = 1/2^-$ and $3/2^-$, both with angular momentum $\mathrm{L}=1$.
These baryons can be neutral (valence quark content {\cPqu-\cPqs-\cPqb}) or negatively
charged ({\cPqd-\cPqs-\cPqb}).
At the Tevatron, baryons with masses and decay modes consistent with the theoretical predictions for the ground
state \PGXb baryons have been observed~\cite{xibd0,xibcdf,xib0}, although their quantum numbers have not yet been
established.
The allowed decays of the experimentally missing \PGXb states should be analogous to the charmed sector~\cite{xic1,xic2,xic3}.
In addition, theoretical calculations~\cite{hb2,hb3,hb4,hb5,hb6} predict the mass difference between the \PGXprb and \PGXb to be
smaller than the mass of the pion, in which case the strong decay $\PGXprb \to \PGXb \Pgp$ is kinematically forbidden. The mass
difference between the \PGXstb and \PGXb, however, is expected to be large enough to allow such a decay.

This Letter presents a search for the decay $\PGXstbz \to \PGXbm \Pgpp$, with $\PGXbm \to \JPsi \, \PgXm$, $\JPsi \to \Pgmp\Pgmm$,
$\PgXm \to\PGLz \Pgpm$, and $\PGLz \to \Pp \Pgpm$. Charge conjugate states are implied throughout.
The reconstruction of such decays involves the presence of three secondary vertices, where the \PGXbm, \PgXm, and \PGLz decay,
which are well separated from the primary interaction vertex.
The analysis is based on a data sample of {pp} collisions at $\sqrt{s}=7$\TeV, collected in 2011 by the Compact Muon Solenoid (CMS) experiment at the Large Hadron Collider (LHC), corresponding to an integrated luminosity of 5.3\fbinv.

The CMS apparatus is described in detail in Ref.~\cite{cms}. Its central feature is a superconducting solenoid, of 6\unit{m} internal diameter, providing a field of 3.8\unit{T}. The main subdetectors used in this analysis are the silicon tracker and the muon systems. The silicon tracker, composed of pixel and strip detector modules, is immersed in the magnetic field, and enables the measurement of charged particle momenta over the pseudorapidity range $|\eta| < 2.5$, where $\eta =-\ln(\tan{\theta}/{2})$ and $\theta$ is the polar angle of the track relative to the counterclockwise beam direction. Muons are identified in the range $|\eta| < 2.4$ using gas-ionization detectors embedded in the steel return yoke of the magnet.

The events used in this analysis were collected using the two-level trigger system of CMS\@.
The first level consists of custom hardware processors and uses information from the muon systems to select
events with two muons. The ``high-level trigger'' processor farm further decreases the event rate
before data storage, requiring two opposite-sign muons
compatible with being the decay products of a \JPsi, either promptly produced or displaced from the primary vertex (PV).
The nonprompt \JPsi trigger requires two opposite-sign muons with an invariant mass within 200\MeV of the
\JPsi mass~\cite{pdg}, single muon transverse momentum $\pt > 3$, 3.5, 4, or 5\GeV, and dimuon vertex fit $\chi^2$
probability larger than 10\% or 15\%. The requirements were made tighter depending on the LHC instantaneous luminosity
so as to limit the trigger rate.
In addition, the dimuon vertex must be separated from the beam line by more than 3 times
the uncertainty on the separation $\sigma_\text{vertex}$, which includes the transverse vertex
resolution (independently estimated for each event) and the beam line position uncertainty.
Finally, the requirement $\cos \alpha> 0.9$ is also applied, where $\alpha$ is the angle, in the plane transverse
to the beam, between the dimuon momentum and the direction from the beam line to the dimuon vertex.
The prompt \JPsi trigger requires two opposite-sign muons with an invariant mass within 250\MeV of the
\JPsi mass~\cite{pdg}, dimuon $\pt > 10$ or 13\GeV (depending on the instantaneous luminosity),
dimuon rapidity $|y| < 1.25$, and dimuon vertex fit $\chi^2$ probability greater than 0.5\%.

The reconstruction of the $\PGXbm \to \JPsi \, \PgXm$ candidates begins by identifying $\JPsi \to\Pgmp\Pgmm$ decays.
The candidate \JPsi mesons are built by combining pairs of oppositely-charged muons (tracks in
the silicon tracker matched to tracks in the muon detectors) satisfying the trigger conditions
and having an invariant mass within 150\MeV of the \JPsi mass.
Candidate \PGLz baryons are reconstructed in decays to a pion ($\Pgp_{\PgL}$) and a proton (\Pp) with opposite charges,
where the higher momentum track is assumed to be the proton. Both tracks are required to have a track fit $\chi^2/\mathrm{ndf} < 5$,
where ndf is the number of degrees of freedom, and at least six hits in the silicon tracker.
The \PGLz candidate vertex must have a fit $\chi^2/\mathrm{ndf} < 7$ and be displaced from the beam line by more than
$10\, \sigma_\text{vertex}$.
Possible contamination from misreconstructed \PKS mesons is removed by requiring that the candidate mass, when assuming
that both tracks are pions, differs from the \PKS mass~\cite{pdg} by more than 20\MeV.
Candidate \PgXm\ baryons are reconstructed by combining a \PGLz candidate with a track ($\Pgp_{\PgX}$) of the same
charge as the $\Pgp_{\PgL}$. A kinematic vertex fit~\cite{kinfit} is performed, assuming that the track is a pion,
and with the \PGLz candidate mass constrained to its world-average value~\cite{pdg}. A veto window of 20\MeV around the
\PgOm~mass~\cite{pdg} is applied when assuming that the $\Pgp_{\PgX}$ candidate is a kaon. The \PgXm\ is then
combined with the \JPsi to form a \PGXbm candidate, with a kinematic vertex fit constraining their masses to the
world-average values~\cite{pdg}.
On average, the events used in this analysis have around eight reconstructed primary vertices (PV).
The PV closest in three dimensions to the \PGXbm trajectory is taken as the one where the \PGXbm
was produced.

The \PGXbm signal selection criteria are chosen by an iterative algorithm that maximizes both the signal yield
and the significance, calculated as $S/\sqrt{S+B}$, where the signal ($S$) and background ($B$) yields are evaluated
in a window of $\pm40$\MeV around the \PGXbm mass, $5790.5\pm 2.7$\MeV~\cite{pdg}. The value of $B$ is linearly
interpolated to this signal region from the two mass sidebands, located between 40 and 100\MeV away from the peak value.
Thirty variables are used to optimize the \PGXbm signal selection. At every iteration, two variables are chosen randomly, where one of them is tightened and the other loosened. If $S$ and the significance increase, the new set of values is accepted. To avoid ending up in a local maximum, an iteration is accepted if the significance decreases by less than a random number uniformly generated between 0 and 10\%.
The variables used for the selection include the \pt of both muons, and of the \JPsi, \Pp, $\Pgp_{\PgL}$,
$\Pgp_{\PgX}$, \PgXm, and \PGXbm. The muons and \PGXbm can have different \pt selections in the
pseudorapidity regions $|\eta| < 1.2$ and $|\eta| > 1.2$, where $|\eta|=1.2$ is the transition between the barrel and endcap detectors.
The algorithm also tunes the requirements on the differences between the reconstructed masses and the world-average masses for the
\PGLz, \PgXm, and \JPsi, with the latter case being different for $|\eta(\JPsi)| < 1.2$ and $|\eta(\JPsi)| > 1.2$. Also, the
total \JPsi pseudorapidity coverage can be adjusted to be smaller than the full $|\eta(\JPsi)| < 2.4$ range.

To select a \PGXbm sample with a good signal-over-background ratio, it is important to reject promptly produced tracks.
This is done via a selection on their transverse impact parameter significances $D_\mathrm{ip}/\sigma_{D\mathrm{ip}}$, where the impact parameter $D_\mathrm{ip}$
is calculated with respect to the beam line and $\sigma_{D\mathrm{ip}}$ is its uncertainty.
This procedure is applied to the \Pp, $\Pgp_{\PgL}$, and $\Pgp_{\PgX}$ tracks, as well as to the \PGLz
and \PgXm\ trajectories.
Long-lived particles (\PGLz, \PgXm, and \PGXbm) are selected by requirements on their transverse decay lengths and
on the significance $L_{xy}/\sigma_{Lxy}$ of the transverse distance between their production and decay vertices $L_{xy}$,
where $\sigma_{Lxy}$ is the uncertainty of $L_{xy}$.
Minimal fit confidence levels %(CL)
are required on the \PGLz, \PgXm, and \PGXbm decay vertices,
while the three-dimensional distance significance %($D_\mathrm{3d}/\sigma_{D3\mathrm{d}}$)
must be smaller than the optimized thresholds for the distances between
the \PgXm\ trajectory and the \JPsi decay vertex %($D_\mathrm{3d}({\PgXm}-\JPsi)$)
, as well as between the \PGXbm trajectory and the chosen PV %($D_\mathrm{3d}({\PGXbm}-\mathrm{PV})$)
.

\begin{figure}[htb]
  \begin{center}
    \includegraphics[width=7cm]{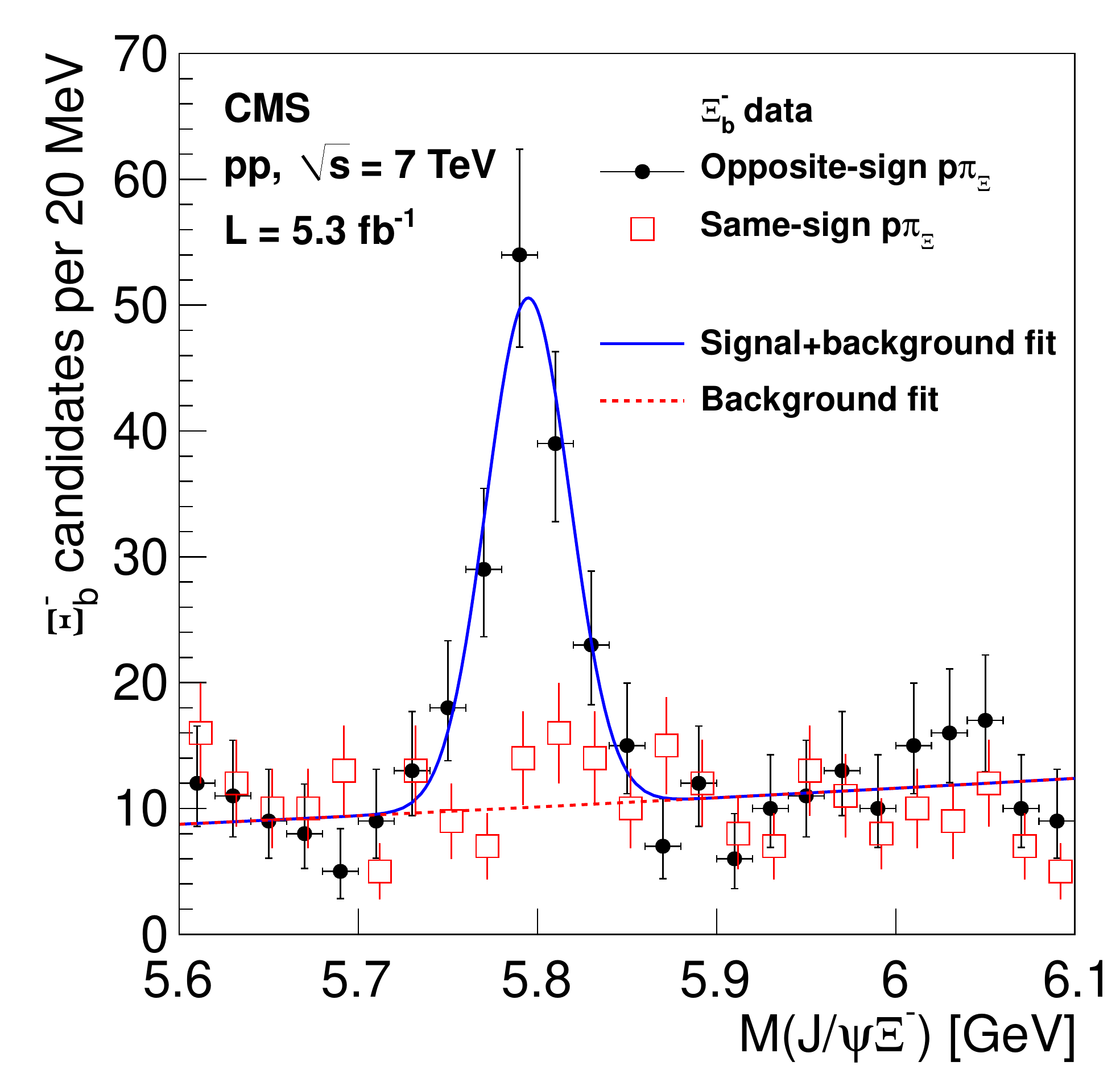}
    \caption{Invariant-mass distributions of the selected \PGXbm candidates (closed circles) and of the $\JPsi \, \PgXm$ pairs where the proton and
    $\Pgp_{\PgX}$ have the same charge (open squares).
    The blue solid curve represents the result of the signal-plus-background fit to the \PGXbm distribution,
    while the red dashed line shows the corresponding background component.}
    \label{fig:xib}
  \end{center}
\end{figure}

The $\JPsi \, \PgXm$ invariant-mass distribution for the \PGXbm candidates passing the selection criteria is shown in Fig.~\ref{fig:xib},
together with the result of a fit with a Gaussian function representing the signal plus a second-order polynomial representing the background,
which gives a signal yield of $108 \pm 14$ events.
The same figure also displays the invariant-mass distribution for background events, in which the
$\Pgp_{\PgX}$ and proton have the same charge, giving further evidence that the observed peak corresponds to a real \PGXbm signal.
The \PGXbm mass extracted from the fit is $5795.0 \pm 3.1 \, \text{(stat.)}$\MeV, in good agreement with the world-average value~\cite{pdg}.
The corresponding mass resolution is $23.7 \pm 3.2 \, \text{(stat.)}$\MeV, in agreement with the value $22.5 \pm 4.7$\MeV,
obtained from a detailed Monte Carlo (MC) simulation of the CMS detector response,
using {\PYTHIA~6.409}~\cite{pythia}, \textsc{EvtGen}~\cite{evtgen}, and \GEANTfour~\cite{geant4}.

To search for \PGXstbz baryons, the \PGXbm candidates with a mass within 2.5 standard deviations of the fitted peak value
are combined with tracks, assumed to be pions,
with a charge opposite to the $\Pgp_{\PgX}$ charge (opposite-sign pairs)
and
coming from the selected PV, with a significance less than 3 standard deviations
on the distance between the track trajectory and the PV\@.
Other quality requirements applied to the tracks are $\pt > 0.25$\GeV, at least two hits in the
silicon pixel layers, at least five hits in the entire tracker, and a track fit $\chi^2/\mathrm{ndf} < 2.5$.

\begin{figure}[htb]
  \begin{center}
    \includegraphics[width=7cm]{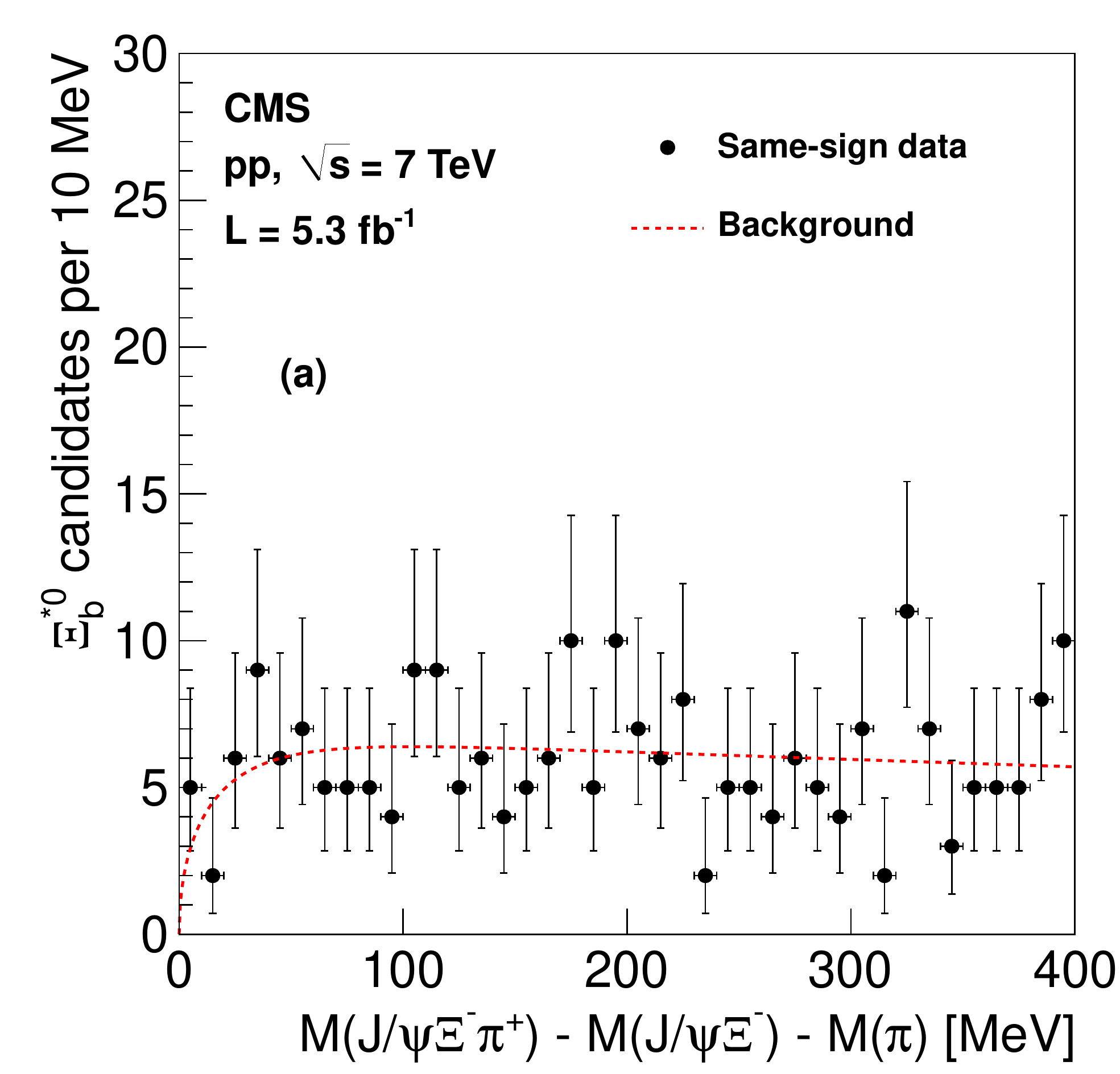}
    \includegraphics[width=7cm]{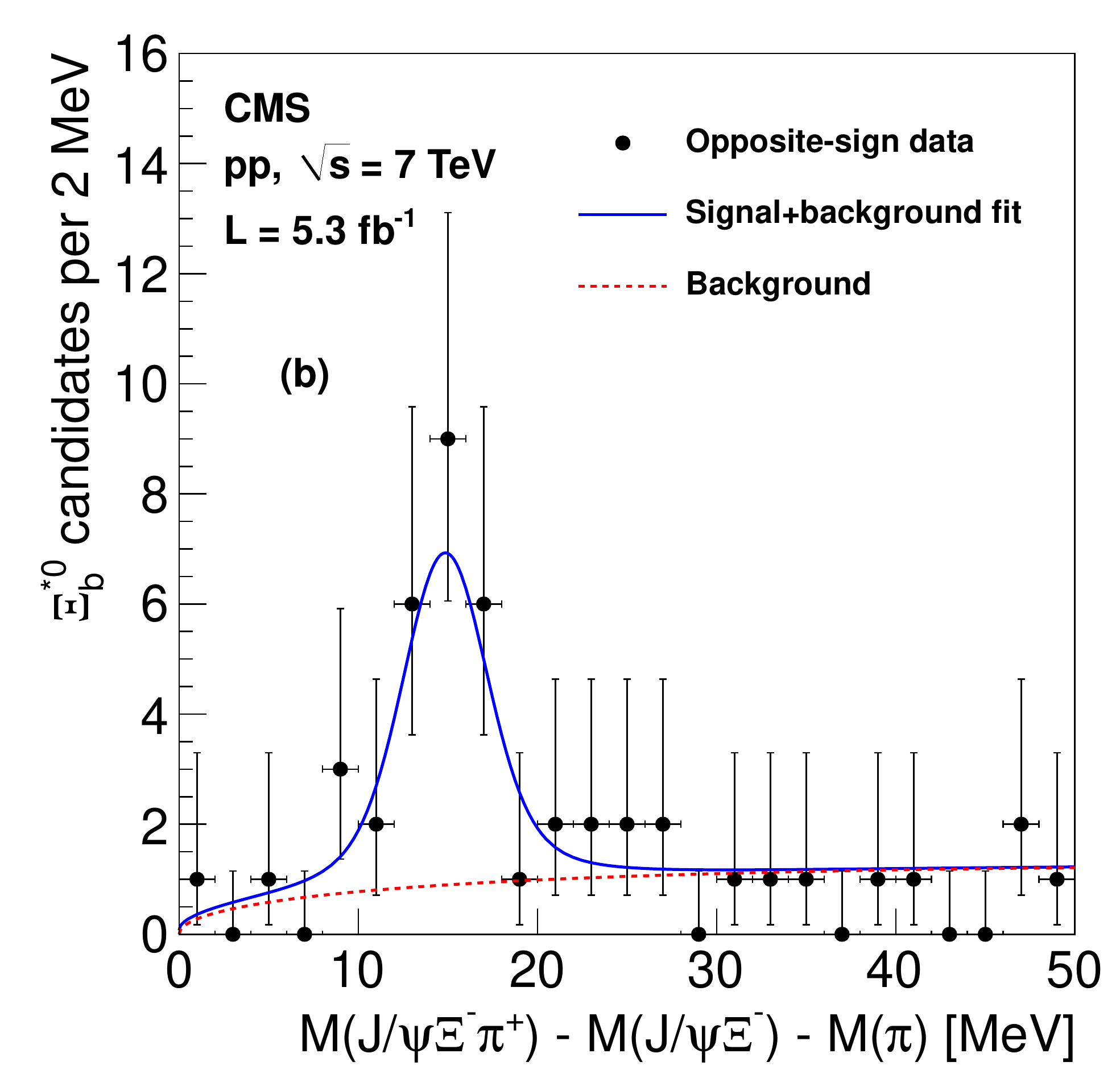}
    \caption{(a) Same-sign $Q$ distribution (closed circles) and result of a fit with the background model (red dashed curve) in the range $0 < Q < 400$\MeV.
    (b) Opposite-sign $Q$ distribution (closed circles) in the $0 < Q < 50$\MeV range, along with the result of the signal-plus-background fit (blue solid curve);
    the background term is also shown (red dashed curve).}
    \label{fig:xib0}
  \end{center}
\end{figure}

The \PGXstbz search uses the mass difference $Q$ between the measured $\JPsi \, \PgXm \Pgpp$ invariant mass and the sum of the masses
of the decay products, $Q = M(\JPsi \PgXm \Pgpp) - M(\JPsi \PgXm) - M(\Pgp)$, where $M(\Pgp)$ is the charged-pion mass~\cite{pdg}.
The search for new resonances in the $Q$ distribution requires a reliable
background
shape. A background model is built using candidates where the prompt pion and the \PGXbm have the same
charge (same-sign pairs), given that the background is expected to be dominated
by combinatorial sources, as checked by MC studies.
The measured momentum distributions of \PGXbm candidates and same-sign pions ($p(\PGXb)$, $p(\Pgp)$), together with the distribution of the angle between them ($\alpha$), are used
to randomly generate uncorrelated values for $p(\PGXb)$, $p(\Pgp)$, and $\alpha$.
Given the limited statistical precision of the \PGXbm momentum distribution, the corresponding random numbers are generated from
a parametrized version using the fit function $f_{\PGXbm}(p) = p^{k_1} \, \re^{-k_2 p}$, where $k_i$ are free parameters.
The three random values are then combined to calculate a $Q$ value for predicting the combinatorial
background distribution.
One hundred million $Q$ values are generated in this way and the resulting distribution is fitted to the function
$Q^{c_1} \, (\re^{-c_2 Q} + \re^{-c_3 Q} + \re^{-c_4 Q})$, where $c_i$ are free parameters.
Figure~\ref{fig:xib0} (a) compares the $Q$ distribution of the same-sign \PGXstbz candidates with the predicted
background shape.
Alternative functional forms of $f_{\PGXbm}(p)$ are used to
estimate the systematic uncertainty associated with this method, which contributes to the determination of the background parameters.

The measured opposite-sign $Q$ distribution is displayed in Fig.~\ref{fig:xib0} (b) for the range 0--50\MeV.
The 21 events observed in the region $12 < Q < 18$\MeV represent a clear excess with respect to the expected
background yield of $3.0 \pm 1.4$ events, evaluated by integrating the background function in this $Q$ window.
An unbinned maximum-likelihood fit is performed to the opposite-sign $Q$ distribution with a
Breit--Wigner distribution convolved with a Gaussian function, added to the background function previously described.
The Gaussian resolution of the peak is constrained to $1.91 \pm 0.11$\MeV, as determined in the signal MC simulation, and
the background parameters are allowed to float within their total uncertainties (statistical plus systematic, added in quadrature).
Figure~\ref{fig:xib0} (b) also shows the result of the fit. A peak is clearly visible above the background continuum.
The fitted parameters of the peak are $Q = 14.84 \pm 0.74 \, {\rm (stat.)}$\MeV and Breit--Wigner width $\Gamma = 2.1 \pm 1.7 \, {\rm (stat.)}$\MeV.
The fitted Breit--Wigner width agrees with $\Gamma = 0.51 \pm 0.16$\MeV, the value obtained following Eq.~(102) of Ref.~\cite{hb}, based on lattice quantum chromodynamic calculations.

To evaluate the significance of the signal, the likelihood $L_{s+b}$ of the signal-plus-background fit is determined.
The fit is then repeated using the background-only model to obtain a new likelihood $L_b$.
The parameters obtained from the background fit are allowed to float so that their uncertainties and correlations contribute
to the calculation of the significance.
The logarithmic likelihood ratio $\sqrt{\ln(L_{s+b}/L_b)}$ would indicate a statistical significance of 6.9 standard deviations ($\sigma$), corresponding to a
probability of $2.5 \times 10^{-12}$ for a background fluctuation of this significance or more to be observed.
The significance remains the same if the fit is repeated using the theoretically expected width, allowing it to vary within the range of the theoretical uncertainty.
The signal significance is also evaluated by generating pseudoexperiments in which the background distribution is varied within
its statistical uncertainty, and determining the background fluctuation probability (``p-value") as the number of experiments
that give a fit with the same significance or higher than in the data.
The ``look-elsewhere effect''~\cite{lee} is assessed by searching for a peak, of width $\Gamma$ between 0 and 25\MeV,
in the extended mass range $0 < Q < 50$\MeV, where the
\PGXstbz is theoretically expected.
The resulting background fluctuation probability is $p = 1.3 \times 10^{-8}$, which corresponds to a $5.7\sigma$ equivalent Gaussian significance.
If the search range is further extended to $0 < Q < 400$\MeV, the equivalent Gaussian significance becomes $5.3\sigma$.

This analysis has been repeated using simulated \PBp, \PBz, \PBs, and \PgLb samples, obtained by the
detailed MC simulation of the CMS detector already mentioned. The samples contain events in which the \cPqb hadron is forced to
decay to a \JPsi, which decays to $\Pgmp\Pgmm$.
No evidence of peaks due to partially reconstructed b-hadron decays is observed in these samples.
The opposite-sign $Q$ distribution obtained using \PGXbm candidates from the lower- and higher-mass sidebands of the
signal peak also shows no excess, indicating that the observed peak is not caused by fake \PGXbm candidates.

The systematic uncertainty on the measured $Q$ value is evaluated through a detailed MC simulation.
The reconstructed $Q$ value in MC simulations is measured to be $0.23 \pm 0.10$\MeV above the generated value.
This is consistent with the observation that the measured \PGLz and \PgXm\ masses are
$0.16 \pm 0.05$ and $0.18 \pm 0.14$\MeV, respectively, above their world averages.
The sum in quadrature of the shift and its statistical uncertainty, 0.25\MeV, is considered as the systematic
uncertainty due to this effect.
As an extreme fitting scenario, a flat function is used for the background shape, leading to a $Q$ value
0.12\MeV higher than the value measured with the nominal background model. Adding in quadrature
this uncertainty with the previous one results in a total $Q$ systematic uncertainty of 0.28\MeV.

The observation of this resonance, corresponding to the one observed
in the charm sector~\cite{xic1}, and its mass measurement add valuable
information to the understanding of the interactions between quarks
within a baryon.

In summary, a new \PGXb baryon has been observed in {\Pp\Pp} collisions at $\sqrt{s} = 7$\TeV, using data
collected by the CMS experiment, corresponding to an integrated luminosity of 5.3\fbinv. The signal is observed with
a significance exceeding 5 standard deviations.
The measured $Q = M(\JPsi \PgXm \Pgpp) - M(\JPsi \PgXm) - M(\Pgp)$ value is $14.84 \pm 0.74\, \text{(stat.)} \pm 0.28\, \text{(syst.)}$\MeV.
Given the charged-pion and \PGXbm masses~\cite{pdg}, the resulting \cPqb-baryon mass is
$5945.0 \pm 0.7\, \text{(stat.)} \pm 0.3\, \text{(syst.)} \pm 2.7\, \mathrm{(PDG)}\MeV$, where the last uncertainty reflects the present accuracy of the \PGXbm mass from the Particle Data Group~\cite{pdg}.
While the width of the new baryon is not measured with good statistical precision, it is compatible with theoretical
expectations~\cite{hb}.
Given its measured mass and decay mode, the new baryon is likely to be the \PGXstbz, with J$^\mathrm{P}=3/2^+$.

\bigskip

We congratulate our colleagues in the CERN accelerator departments for the excellent performance of the LHC machine. We thank the technical and administrative staff at CERN and other CMS institutes, and acknowledge support from: FMSR (Austria); FNRS and FWO (Belgium); CNPq, CAPES, FAPERJ, and FAPESP (Brazil); MES (Bulgaria); CERN; CAS, MoST, and NSFC (China); COLCIENCIAS (Colombia); MSES (Croatia); RPF (Cyprus); MoER, SF0690030s09 and ERDF (Estonia); Academy of Finland, MEC, and HIP (Finland); CEA and CNRS/IN2P3 (France); BMBF, DFG, and HGF (Germany); GSRT (Greece); OTKA and NKTH (Hungary); DAE and DST (India); IPM (Iran); SFI (Ireland); INFN (Italy); NRF and WCU (Korea); LAS (Lithuania); CINVESTAV, CONACYT, SEP, and UASLP-FAI (Mexico); MSI (New Zealand); PAEC (Pakistan); MSHE and NSC (Poland); FCT (Portugal); JINR (Armenia, Belarus, Georgia, Ukraine, Uzbekistan); MON, RosAtom, RAS and RFBR (Russia); MSTD (Serbia); MICINN and CPAN (Spain); Swiss Funding Agencies (Switzerland); NSC (Taipei); TUBITAK and TAEK (Turkey); STFC (United Kingdom); DOE and NSF (USA).

\bibliography{auto_generated}   % will be created by the tdr script.

\cleardoublepage \appendix\section{The CMS Collaboration \label{app:collab}}\begin{sloppypar}\hyphenpenalty=5000\widowpenalty=500\clubpenalty=5000\textbf{Yerevan Physics Institute,  Yerevan,  Armenia}\\*[0pt]
S.~Chatrchyan, V.~Khachatryan, A.M.~Sirunyan, A.~Tumasyan
\vskip\cmsinstskip
\textbf{Institut f\"{u}r Hochenergiephysik der OeAW,  Wien,  Austria}\\*[0pt]
W.~Adam, T.~Bergauer, M.~Dragicevic, J.~Er\"{o}, C.~Fabjan\cmsAuthorMark{1}, M.~Friedl, R.~Fr\"{u}hwirth\cmsAuthorMark{1}, V.M.~Ghete, J.~Hammer, N.~H\"{o}rmann, J.~Hrubec, M.~Jeitler\cmsAuthorMark{1}, W.~Kiesenhofer, V.~Kn\"{u}nz, M.~Krammer\cmsAuthorMark{1}, D.~Liko, I.~Mikulec, M.~Pernicka$^{\textrm{\dag}}$, B.~Rahbaran, C.~Rohringer, H.~Rohringer, R.~Sch\"{o}fbeck, J.~Strauss, A.~Taurok, P.~Wagner, W.~Waltenberger, G.~Walzel, E.~Widl, C.-E.~Wulz\cmsAuthorMark{1}
% ...
\vskip\cmsinstskip
\dag:~Deceased\\
1:~~Also at Vienna University of Technology, Vienna, Austria\\
% ...

\end{sloppypar}
\end{document}